\PassOptionsToPackage{usenames,dvipsnames}{xcolor}
\documentclass[runningheads]{llncs}
\usepackage{graphicx}

\usepackage{xcolor}
\usepackage[inline]{enumitem}
\usepackage{relsize}

\usepackage[square,comma,numbers,sort&compress]{natbib}

\newcommand{\termdefn}[1]{\emph{#1}}
\newcommand{\ie}{i.e., }

\begin{document}
\title{
Adaptive parallelization of multi-agent simulations with localized dynamics}
\titlerunning{Unchained computation} 
%
\author{Alexandru-Ionu\c{t} B\u{a}beanu 
\and
Tatiana Filatova 
\and
Jan H.\@ Kwakkel 
\and
\mbox{Neil Yorke-Smith} 
}
\authorrunning{A.I. B\u{a}beanu et al.}
%
\institute{Delft University of Technology, The Netherlands.  \email{\{A.I.Babeanu,T.Filatova,J.H.Kwakkel,N.Yorke-Smith\}@tudelft.nl}}
\maketitle              
\begin{abstract}
Agent-based modelling constitutes a versatile approach to representing and simulating complex systems. 
Studying large-scale systems is challenging because of the computational time required for the simulation runs: scaling is at least linear in system size (number of agents).
Given the inherently modular nature of MABSs, parallel computing is a natural approach to overcoming this challenge.
However, because of the shared information and communication between agents, parellelization is not simple.
We present a protocol for shared-memory, parallel execution of MABSs.
This approach is useful for models that can be formulated in terms of sequential computations, and that involve updates that are localized, in the sense of involving small numbers of agents.
The protocol has a bottom-up and asynchronous nature, allowing it to deal with heterogeneous computation in an adaptive, yet graceful manner. 
We illustrate the potential performance gains on exemplar cultural dynamics and disease spreading MABSs.

\keywords{agent-based models \and large-scale systems \and parallel computing \and scientific simulation \and asynchronous computation}
\end{abstract}


\section{Introduction}
\label{sec:intro}

Whereas multi-agent-based simulation constitutes a versatile approach to representing and simulating social, biological and other types of (complex) systems, the use of MABS to study large-scale systems is challenging due to the computational time required.
Scaling a MABS is linear or worse in the number of agents.
A natural approach to handling this issue is parallel computing.
Whereas parallelizing a set of independent runs of a simulation is straightforward,
using multiple processing cores to accelerate a \emph{single} simulation run is much more challenging.
Despite the modular nature of MABS, agents (and other simulation components) engage in frequent interactions, so their computations cannot be parallelized in simple ways \citep{Parry2020}.

To tackle this challenge, we present a novel protocol for shared-memory, parallel execution of MABSs.
We exploit the fact that a MABS may be expressed as a sequential chain of computational steps,
and that MABS models often involve updates that are \termdefn{localized}, in the sense of involving small numbers of agents.

Compared to most work on parallelizing MABS, our protocol handles the computation in a profoundly asynchronous manner:
different CPU cores may work on different steps of the simulation at any given instant,
and may handle different agents at different times.
This enables adaptive adjustments to fluctuations due to heterogeneity inherent to the simulated process or the underlying infrastructure.
The protocol also facilitates parallelization of MABS models that are fully sequential in form.
Potential performance gains, which we illustrate using two MABSs (Axelrod-like cultural dynamics and network-based disease spreading), 
are a consequence of computing resources being fully used at every instant,
at the cost of carrying out additional computation.
We provide insights about the associated protocol overhead and guidelines for managing it. 



\section{Related Work}
\label{sec:rw}

The challenge of scaling up MABS has seen recurring efforts in the literature \citep{Parry2020,Axtell2022}.
One line of work exploits the notion of a \termdefn{super-agents}: subsets of agents with some similarity are combined into representative agents \citep{Scheffer1995}.
Along these lines, \citet{Arneth2014} introduce a concept of \termdefn{agent functional types}, which can be seen as super-agents corresponding to different strategies or styles.  
Related to but distinct from (manually-coded) super-agents, \citet{DBLP:conf/mabs/TregubovB20} discuss methods to automatically simplify a MABS, such as by automatic abstraction.
Super-agents have been found to be problematic in spatial settings \citep{Parry2020}.

A second paradigm employs surrogate models that emulate interesting properties of MABS \citep{DBLP:journals/jasss/BroekeVLM21, Lamperti2018}.  
This attracts more attention as capabilities of machine learning techniques increase \citep{Angione2022,DBLP:conf/mabs/LeeuwZS22}, but do not directly address accelerating a given MABS.
Our ambition is to avoid any re-programming or approximation.

A third approach, which is closest to our work, aims to use high-performance computing (HPC) techniques and hardware for running MABS.
Along these lines, \citet{Blandin2017} present a distributed-memory, message-passing approach to parallelizing certain types of agent-based simulations in Python, illustrated with a migration model where it can handle up to 10 billion agents -- while still employing certain model simplifications and abstractions designed to reduce communication between processes. 
\citet{Blythe2018} is another example of an architecture for massively distributed MABS execution, leveraging contemporary multi-processing and load balancing infrastructure, combined with graph-partitioning techniques.
By contrast, \citet{Axtell2016} runs 120 million agents on a single, multi-core, shared memory system, reporting that ``Many HPC architectures turns out to not be useful''. 
Generic computational frameworks have also been developed for agent-based-like simulations in biology, such as those described in \citet{Breitwieser2021} and \citet{Plesser2007}, which allow for jointly using distributed-memory and shared-memory parallelization.

While scale-up already demonstrated by HPC-driven efforts is remarkable, we note that most research to date is subject to a fundamental limitation:
parallelization goes hand in hand with strictly splitting the computation into time steps and updating (a step-dependent subset of) all agents at each step. 
Different subsets of updates within that step are then allocated to available CPU cores, using some workload-balancing strategy.
While some performance may be gained from context-dependent improvements of the allocation strategy,
computing cores/nodes that eventually run out of work may not proceed to the next step until the current step has been completed.
This is at odds with the local nature of many MABS models, allowing agents to update many steps ahead of other agents, as long as the former have the required input at every step. 
Moreover, the per-step splitting approach cannot be applied to models that lack the many-updates-per-step formulation in the first place -- for instance, Axelrod-type models of cultural dynamics \citep{Axelrod1997} and Markov-Chain simulations, which have a sequential, one-update-per-step formulation -- but could still benefit from parallelization. 
We propose a generic way of jointly handling these issues, which at heart is a more flexible approach to handling \termdefn{time} in scientific computing, 
which we present as a protocol for shared memory, adaptive parallelization. 

Our proposed protocol is reminiscent of existing work on (software-based) \termdefn{dynamic scheduling}, going back at least as far as \citet{Blumofe1995}, with a good overview of relevant tools and concepts presented by \citet{Ham2012}.
The openMP library \citep{dagum1998openmp}, which is widely used for parallel HPC in a shared memory context, also incorporates the option for dynamic scheduling.
However, dynamic scheduling focuses on parallelizing computational tasks that are independent of each other, whereas our protocol is designed to gracefully handle precisely such dependence relations between tasks.
Also, our protocol incorporates a clean procedure for interfacing with any given MABS model.

Sensitivity to dependence relations is present in a different approach to adaptive, asynchronous parallelization of MABS, proposed by \citet{DBLP:journals/jpdc/ScheutzS06, DBLP:series/sci/ScheutzH10}.
Their work is complementary to ours in multiple ways:  
\begin{enumerate*}
    \item it uses a distributed memory paradigm with message passing between processes, while our protocol heavily relies on shared memory;
    \item it ensures consistency via \termdefn{proxy entities} and \termdefn{event horizons}, while we ensure it via worker \termdefn{records} and task \termdefn{recipes};
    \item each of their CPUs handles a preallocated subset of agents at all times, while our cores may handle different agents at different times 
    \item it comes with a sophisticated formalism (involving constructs like bodies, sensors and actuators) in terms of which the model needs to be expressed, while our protocol is less involved and `invasive' in that sense;
    \item it is focused on spatial models with moving agents and maximum movement velocities, while our applications do not yet involve moving agents.
\end{enumerate*}
Ideas from the two approaches could be gradually combined in future work.


\section{Methodology: Computational Protocol}
\label{sec:protocol}

The central idea is that a MABS may be conceptualized as a \termdefn{chain}, where each element in the chain is a \termdefn{task}.
Each task is a block of consecutive operations that are carried out at some point during the simulation.  
An operation may consist, for instance, of generating a random number or incrementing a variable.
Normally, these tasks would be evaluated sequentially, based on their order in the chain.
In many cases, however, one may alter the evaluation order without changing the outcome of the simulation: there are many pairs of tasks that are entirely independent of each other, and may thus be evaluated in any order, or in parallel.
For example, two agents whose next decisions are influenced only by their neighbours, and who have no neighbours in common, are independent for their next step.

Our protocol is designed to take advantage of precisely this aspect, 
allowing for a significant amount of parallelization, 
while preserving the evolution of the system. 
This is achieved in an adaptive manner, with multiple computational \termdefn{workers} autonomously iterating the chain and executing those tasks that can be executed without violating model-dependent constraints (manifested as dependence relations between tasks). 
While the amount of parallelization and resulting speed-up is ultimately constrained by the number of available CPU cores (\ie one for each worker--thread pair), there will be stronger bounds determined by the MABS \termdefn{model}, the size of the system and certain choices pertaining to how the protocol is applied.

\begin{figure}[tb]
    \includegraphics[width=\textwidth]{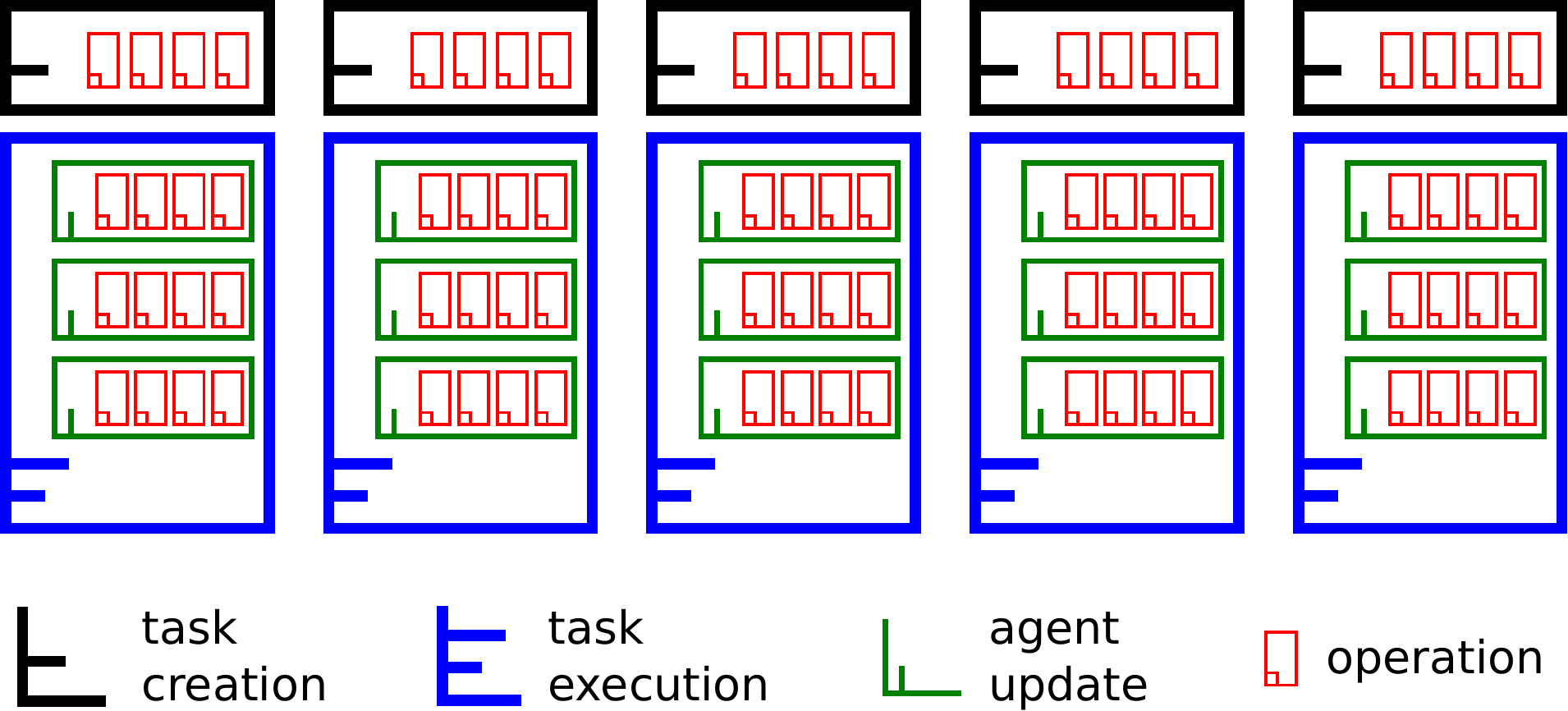}
    \caption{Partitioning the simulation.
    The bottom of the figure is a legend, conveying the meaning of each type of object above, via colour and (partial) shape. 
    Each vertical block corresponds to one task, split between a creation (black) and an execution (blue) part. 
    The small (red) boxes denote the basic operations that make up the simulation.
    Within each task execution, subsets of operations make up agent updates (green).  
    } 
    \label{fig:partition}
\end{figure}

\subsection{Task-based representation}
\label{sec:task-based}

Fig.~\ref{fig:partition} gives an intuitive overview of how the protocol effectively partitions the set of operations that make up a simulation. 
Each resulting subset of operations is associated to one task, with important constraints in place: 
for any two distinct operations $\theta$ and $\rho$ associated to task $\alpha$, 
any operation $\phi$ that depends on $\theta$ and that $\rho$ depends on must also be associated to $\alpha$.
A simulation is mathematically equivalent to the sequence of operations obtained by concatenating all tasks, with an ordering of updates that is consistent with all dependence relations between them.
In the more specific context of a MABS, at the level of each task, many operations are typically grouped into \termdefn{updates}, with each update capturing a change in the state of one or several agents. 

Every task is first \termdefn{created} and added to the end of the chain. 
Later, it is \termdefn{executed} (the entailed operations are carried out) and finally erased from the chain. 
Creating the task already carries out part of the associated computation,
while executing the task carries out the remaining part, 
based on information produced during creation and stored with the task. 
The split between creation and execution is also constrained:
if operation $\rho$ is included in the creation part, any operation $\theta$ on which $\rho$ depends must also be included, assuming that $\theta$ is associated to the same task to start with.
Still, there may be significant freedom for deciding how much computation is carried out during creation vs execution. 

\subsection{Dependence relations between tasks}
\label{sec:dependence-relations}

We assume there is a finite set of variables $V$ that can be repeatedly altered during the simulation.
The execution part of each task $\alpha$ depends on an \termdefn{input subset} of variables $I(\alpha)$ and modifies another, \termdefn{output subset} of variables $O(\alpha)$; the two $\alpha$-subsets may have an arbitrary amount of overlap.
If $\alpha$ precedes another task $\beta$ in the chain (at any separation) and $O(\alpha) \cap I(\beta) \neq \emptyset$, there is a \termdefn{dependence} (or causal link) $\alpha \prec \beta$ between the two:
the execution of an arbitrary $\beta$ can only start after having completed the execution of all $\alpha$ with $\alpha \prec \beta$.
On the other hand, if two tasks $\alpha$ and $\beta$ have no direct or indirect dependence relation, then $\alpha$ and $\beta$ commute: they can be executed in any order, as well as in parallel (to an arbitrary extent).

To be precise, $I(\alpha)$ and $O(\alpha)$ above are defined in a maximally conservative way: they are the subsets of variables that may conceivably influence and, respectively, be influenced by the computation encoded by $\alpha$.
In practice, the subsets of variables actually used/modified will typically be smaller, due to decisions made during the execution of $\alpha$ (this does not lead to redundant causal links: not changing something that could have been changed is also information).
Note that, while conceptually useful for the explanations above, our protocol does not explicitly compute any input or output subsets when inferring dependence relations. Instead, it uses only necessary information (like agent ids) previously recorded onto the chain and workers in a dynamic fashion.

\subsection{Worker-chain dynamics}
\label{sec:worker-chain-dynamics}

Adaptive parallelization is achieved using multiple workers that operate on the chain in a simultaneous but asynchronous manner.
Each worker is assigned to a dedicated thread.
Workers have the freedom to execute tasks as soon as possible, provided that dependence relations between tasks are not violated, that race conditions are absent (since everything takes place within one program run with a fully-shared memory pool), and that the (unavoidable) use of mutex locks does not lead to deadlock situations.
This requires a self-consistent set of rules specifying the joint dynamics of workers and chain: a \termdefn{workflow}.
We focus here on a relatively simple formulation, which we use for the experiments in Sec.~\ref{sec:experimental-evaluation}. 

The chain is implemented as a bidirectional, linked list of tasks, with each task holding pointers to the previous and next tasks, so that erasing tasks inside the chain is cheap and clean. 
At any point in time, every worker is either waiting to enter the chain, or is located at a given task in the chain, with different workers attached to different tasks.
Every worker iterates the chain from start to end.
Upon reaching task $\beta$, worker $w$ is able to infer whether $\beta$ depends on any task $\alpha$ that has already been encountered. 
At this point:
\begin{itemize}
    \item if $\beta$ depends on a previous task, or if the task is already being executed by another worker $u$, the worker $w$ refrains from executing $\beta$; instead, it collects the information describing $\beta$ and integrates it with already collected information from previously-encountered tasks, which it can use to infer dependence relations for subsequent tasks; then $w$ attempts to move to the next task
    \item otherwise, $w$ proceeds to execute $\beta$, erase $\beta$ from the chain and return to the start of the chain (while discarding all information about previously encountered task)
\end{itemize}
At most one task is created at any instant and added to the end of the chain, by a worker attempting to leave the last element.  It does so by invoking a global, model-specific routine.

It is crucial that worker $w$ cannot move to a task $\gamma$ where another worker $v$ is already located, unless the latter is already executing $\gamma$. 
The implied waiting of one worker behind the other is handled by a dedicated mutex lock attached to each task in the chain.

Two further mutex locks, namely an \termdefn{enter-lock} and an \termdefn{erase-lock} are attached to the chain itself.
Specifically, the enter-lock treats the case of a new task being created when the chain is empty (with no task and no associated locks present).
This situation is guaranteed to occur at the start of the process and is likely to occur many times afterwards. 
Hence, the new task is created by the first worker entering the chain, while holding  the enter-lock.
Second, the erase-lock ensures that at most one task is being erased at any given point in time (by the worker that just executed it), thus avoiding inconsistencies related to simultaneously erasing consecutive tasks.

We use the term \termdefn{cycle} to refer to one round of chain exploration by a given worker, between two consecutive returns to the start of the chain.
A cycle ends either when the worker has just erased a task, or when it has reached the end of the chain and cannot create a new task. 
The latter may happen because 1) all desired tasks have already been generated for the given simulation; 
or 2) the worker has reached the maximum number of tasks that it is allowed to generate per cycle.
The latter tasks-per-cycle limit avoids (unlikely) situations when a worker at the end of the chain indefinitely creates new tasks that cannot be executed because of being dependent on previously-encountered tasks, thus extending the chain to unreasonable lengths and potentially increasing the exploration time for all other workers. 
This upper bound is controlled by a parameter.

\subsection{Choices in applying the protocol}
\label{sec:choices}

We briefly note three conceptual choices being made when applying the protocol to a given MABS model. 
\begin{enumerate*}
\item \termdefn{Chain granularity} refers to the amount of computation allocated per task.
\item \termdefn{Task depth} defines the split between task creation and execution.
\item \termdefn{Workflow parameters} are, notably, $n$, the number of workers, and $C$, the maximum number of created tasks per cycle.
\end{enumerate*}

\subsection{Model-specific implementation}
\label{sec:model-specific}

The protocol involves a significant amount of functionality that is independent of the MABS model to which it is applied. 
Around this model-independent functionality we created a dedicated software framework (in due course available open source) allowing one to `plug in' any MABS of interest. 
To do this, one implements a model-specific interface, according to a set of conventions that allow for the intended communication between the model and the workflow.
The framework itself is implemented in C+$\!$+.

The interface can be understood in terms of two generic concepts:  
\begin{enumerate*}
    \item \termdefn{recipe}: model-side counterpart of the task;
    \item \termdefn{record}: model-side counterpart of the worker.
\end{enumerate*}
These concepts capture model-specific data-structures and functionality. 
The recipe specifies the kind of information the task holds after its creation, and how that is to be used for its execution. 
The record specifies the kind of information the worker needs to hold, in order to recognize whether the task at hand is dependent on any task previously encountered, along with the procedure for carrying out this assessment. 
It also specifies how the information in the task at hand is integrated with that already held by the worker (if needed) and how the worker-held information is initialized and reset at the start of a new cycle. 


\section{Experimental Evaluation}
\label{sec:experimental-evaluation}

At this first stage of our research, the purpose of experimentation is to establish the extent to which our protocol exhibits, for a given MABS simulation, the expected speed increase as more workers are added.
To achieve this, we measure, for different values of $n \in \{1 \ldots 5\}$, the total simulation time $T$ as a function of a task size proxy $s$.
For small $s$ little speed-up is expected.
We keep $C = 6$ fixed, since separate experimentation showed its effect to be negligible.
For each combination of $s$ and $n$, $T$ is averaged over 5 simulation instances with different starting seeds (and initial states, whose generation does not contribute to $T$).

Two experiments are reported, for two associated MABS models formulated according to different paradigms: a model with sequential interactions involving random pairs of agents in Sec.~\ref{sec:cultural-dynamics}; a model where all agents are updated at each step, conditionally on nearest-neighbours' states during the previous step, in Sec.~\ref{sec:disease-spreading}. The former is mostly known for illustrating a patter-formation mechanism, while the latter is also known for its real-world, predictive potential.
The two experiments also differ in terms of details related to formulating $s$ and using the protocol, as explained further below.

\subsection{Cultural dynamics}
\label{sec:cultural-dynamics}

First we apply our protocol to an Axelrod-type model of cultural dynamics \citep{Axelrod1997}, following the specifications in \citet{Babeanu2018}.
The $N$ agents are all connected to each other, and each is equipped with a sequence of traits, each expressed with respect to one of the $F$ cultural features.
At each step, two agents are chosen at random to interact, with preassigned roles: 
a \termdefn{source} and a \termdefn{target}.
The target may then change one trait, based on a probabilistic procedure that depends on all traits of both agents, which is intended to mimic social influence.

On the protocol side, the granularity is so that each task captures one pairwise interaction. 
The depth is so that creation handles the random selection of the interacting pair, leaving the bulk of the interaction to the execution.
Hence, the recipe holds the two agents' identifiers, while the record specifies that a task at hand is considered dependent if either the source or the target agent was a target in any task previously encountered by the worker.  

\begin{figure}[tbp]
    \includegraphics[width=\textwidth]{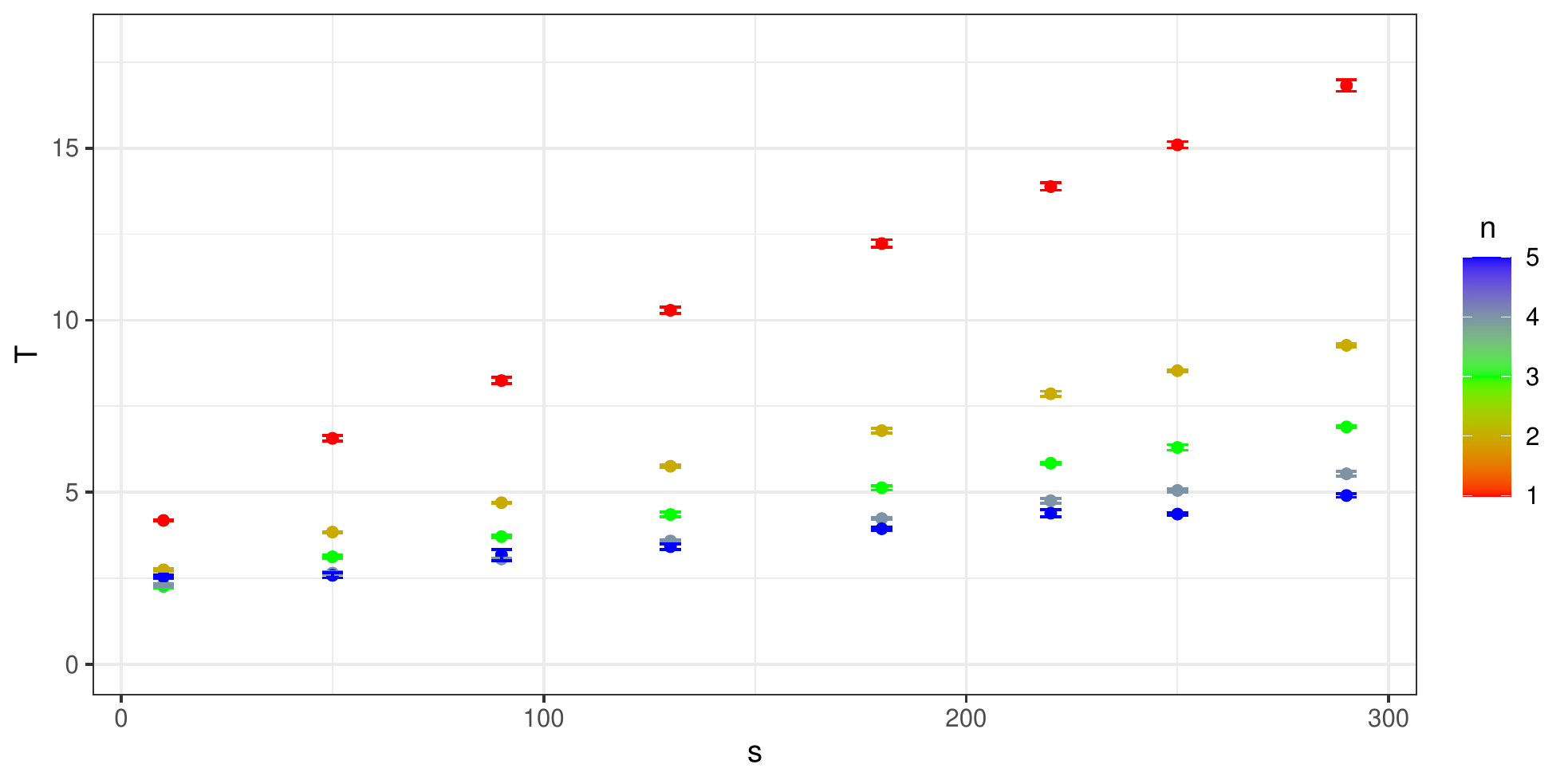}
    \caption{Running cultural dynamics with our protocol.
Showing the average simulation time (vertical axis) as a function of task size proxy (number of features $F$, along horizontal axis), for different numbers of workers (legend). Vertical error bars show the standard mean error range based on five simulation instances. 
} \label{fig:cultural-dynamics-results}
\end{figure}

Fig.~\ref{fig:cultural-dynamics-results} shows the results of the experiment, where $s$ is defined as $F$, since the bulk of one interaction is built around an iteration over all features.
We see that simulation time $T$ increases with task size $s$, independently of $n$, which is in agreement with the fact that the total number of created tasks is fixed. 
Moreover, we see a clear trend of $T$ decreasing with $n$ for a fixed $s$, which confirms the anticipated performance increase as more workers are added.
While the magnitude of this effect (in terms of the ratios between the $T$ values of different $n$ curves) increases with $s$, we also observe a saturation effect: for $s < 150$, it does not help to have more than $n=4$ workers, while for much lower $s$ it does not help (and it might actually harm) to have more than $n=3$ workers.
All this is consistent with the intuition that, as task size increases, the protocol overhead becomes less important. 
This experiment involved $2\times10^6$ steps per run, with the values of the remaining model parameters fixed as: $N = 10^4 $, $q = 3$ (number of possible traits per feature), $\omega = 0.95$ (bounded-confidence threshold).

\subsection{Disease spreading}
\label{sec:disease-spreading}

Second we apply the protocol to a simplified SIR-type epidemic model, which captures essential aspects of predictive MABSs of contagious disease spreading, such as those used in \citet{Gomez2021, Gemmetto2014} for real-world scenarios. 
Specifically, $N$ agents reside at the nodes of a fixed graph with constant degree $k$ and a ring-like structure.
Each agent may be in one of three possible states: susceptible (S), infected (I) or recovered (R).
At each step, for each agent, a transition may occur between two consecutive states, according to simple probabilistic rules controlled by parameters $p_{\mathrm{SI}}, p_{\mathrm{IR}}, p_{\mathrm{RS}} \in (0,1)$.
The transition from S to I also depends on the fraction of infected connections. 

Using the protocol involves additional subtleties compared to the first experiment.
First, there are two types of tasks in the chain: the first type pertains to agents computing their new states; the second type pertains to agents replacing their current states with the new states. 
Second, each task handles a subset of agents, using a partition of the system into equal subsets, which is fixed throughout the simulation.
The aggregation level inherent to this partition specifies the chain's granularity. 
In terms of depth, each creation produces a recipe containing the agent subset identifier, along with a binary flag indicating the task's type, leaving the actual computations for the execution.
The record specifies that: a second-type task should not be executed before a previously-encountered first-type task handling the same agent subset; a first-type task should not be executed before a previously-encountered second-type task handling the same or a connected agent subset. 
Connections between agent subsets are encoded in an aggregate graph computed once (just after generating the initial state); this computation contributes to the measured $T$ (unlike generating the initial state). 

\begin{figure}[tbp]
    \includegraphics[width=\textwidth]{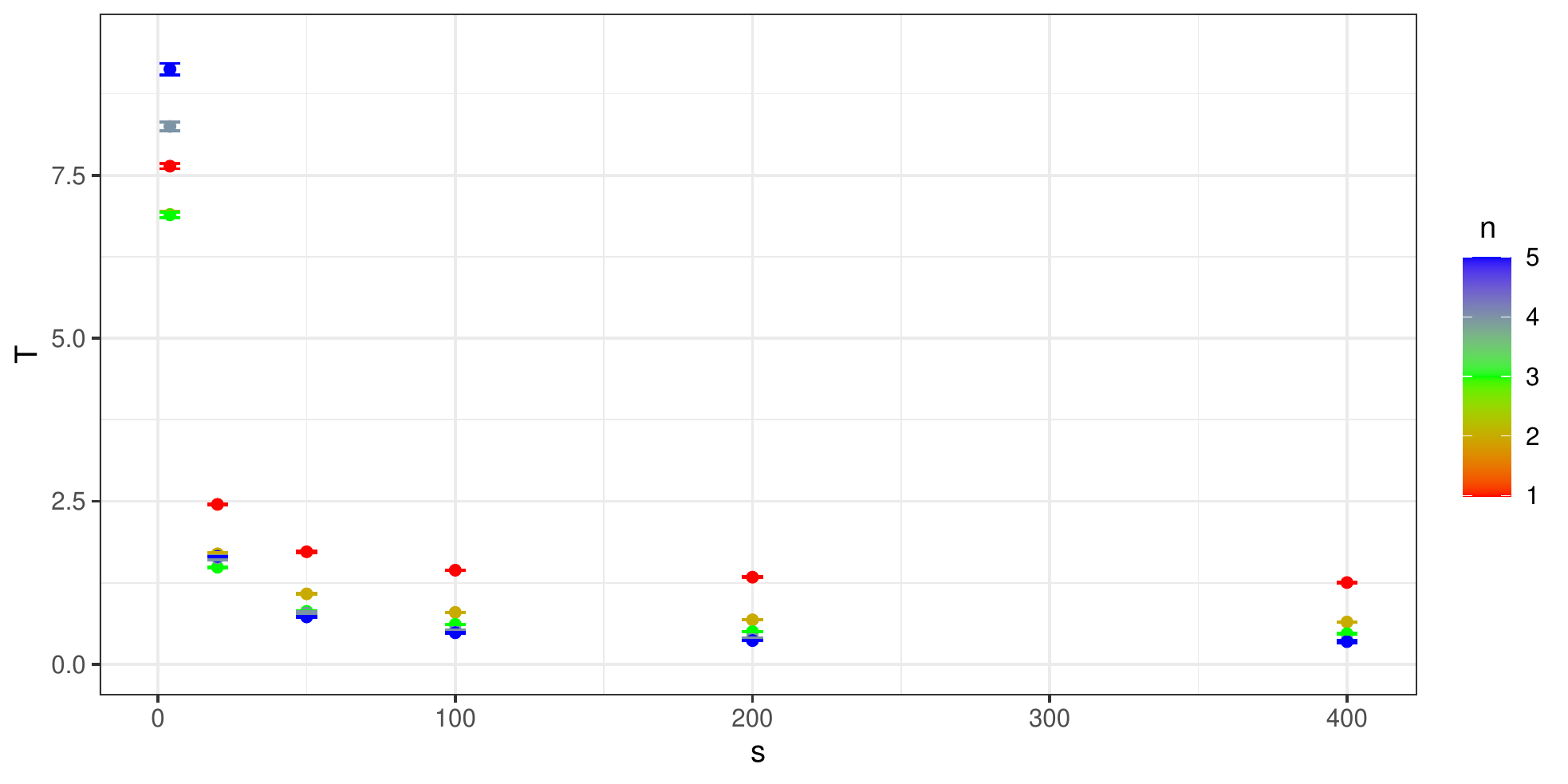}
    \caption{Running disease spreading with our protocol.
Showing the average simulation time (vertical axis) as a function of task size proxy (number of agent updates per task, along horizontal axis), for different numbers of workers (legend). Vertical error bars show the standard mean error range based on five simulation instances.
} \label{fig:disease-spreading-results}
\end{figure}

Fig.~\ref{fig:disease-spreading-results} shows the results of the experiment.
Unlike Sec.~\ref{sec:cultural-dynamics}, there is no model parameter that can serve as $s$. 
Instead, we define $s$ as the size of the agent subsets, which directly controls the task size, as well as the chain's granularity.
A consequence of this, immediately visible in the figure, is that $T$ does not increase with $s$ (as it does in Fig.~\ref{fig:cultural-dynamics-results}), exhibiting instead a sharp decrease, followed by a stabilization: the model-related computation is the same for all $s$, just differently partitioned, so the fine-grained partitioning inherent to $s<50$ becomes very taxing due to protocol overhead, regardless of $n$. 
In the stabilization region we can see the anticipated performance increase with increasing $n$, with saturation reached for $n=4$.
As we go deeper into the low-$s$ region, the saturation is reached for lower $n$, and we clearly see that adding workers may decrease performance.
This experiment involved $3 \times 10^3$ steps per run, with the values of the remaining model parameters fixed as: $N = 4 \times 10^3$, $p_{\mathrm{SI}} = 0.8$, $p_{\mathrm{IR}} = 0.1$, $p_{\mathrm{RS}} = 0.3$, $k = 14$.

\section{Summary and Future Work}
\label{sec:conc}

To unlock the potential of multiple cores for accelerating single runs of a MABS, we presented a novel protocol for shared-memory, parallel execution of multi-agent-based simulations. 
The benefits are seen in experiments on Axelrod-like cultural dynamics and network-based disease spreading simulations.

While our protocol can in principle be applied to any form of computation, it makes sense to use it only if that computation is CPU-bound 
(rather than input--output-bound), as can be the case for MABS with many agents in, for instance, ecologoy \citep{DBLP:journals/envsoft/FilatovaVV11}.
Moreover, we expect an actual speed-up of the computation if, for some choice of chain granularity and task depth, the resulting (directed, acyclic) graph of dependence relations between tasks is sparse enough. 
In other words, a typical task $\alpha$ should only depend on a small part and only affect a small part of the system.
In the context of MABS, `small' typically means that the relevant fractions of input and output variables $I(\alpha)/V$ and $O(\alpha)/V$ scale as $1/N$, where $N$ is the number of agents. 
This is what we mean by \termdefn{localized} dynamics -- the term \termdefn{local} is better suited for more specific MABS where information propagation is constrained by an underlying geometry.
For instance, models involving agents on a lattice that only interact with nearest-neighbours are good candidates for our protocol. 
Even if the model involves global entities coupled to the entire system, the protocol could still be beneficial if the associated interactions are relatively rare, and/or the coupling is only one-way (only influencing or being influenced by the rest of the system). 

Future work will present applications of our protocol to simulations with non-stationary agents or dynamic spatial geometry.
Second, we consider more explicitly using the DAG nature of the computation, 
which could reduce the overhead of the protocol in terms of both memory and CPU usage.
Third, we plan to invoke techniques from \citet{DBLP:series/sci/ScheutzH10} and join the strengths of the approaches.
Beyond MABS, we expect that other types of (non-agent-based) simulations could also benefit from this line of research, including some types of Markov-Chain simulations.

{\smaller
\subsubsection*{Acknowledgements}
This work benefited from interactions with J.\@ 
Decouchant and M.\@ 
Mircea, and was partially supported by the TU Delft Institute for Computational Science and Engineering, and by TAILOR, a project funded by the EU Horizon 2020 programme grant~952215. 
We also thank two anonymous referees for their thoughtful comments on an early version of this article.
}


%
%
%
\renewcommand{\bibsection}{\section*{References}} 
\bibliographystyle{splncs04nat}
\bibliography{main}

\begin{thebibliography}{24}
\providecommand{\natexlab}[1]{#1}
\providecommand{\url}[1]{\texttt{#1}}
\providecommand{\urlprefix}{URL }
\expandafter\ifx\csname urlstyle\endcsname\relax
  \providecommand{\doi}[1]{doi:\discretionary{}{}{}#1}\else
  \providecommand{\doi}{doi:\discretionary{}{}{}\begingroup
  \urlstyle{rm}\Url}\fi

\bibitem[{Angione et~al.(2022)Angione, Silverman, and Yaneske}]{Angione2022}
Angione, C., Silverman, E., Yaneske, E.: Using machine learning as a surrogate
  model for agent-based simulations. PLOS ONE \textbf{17} (2022),
  \doi{10.1371/journal.pone.0263150}

\bibitem[{Arneth et~al.(2014)Arneth, Brown, and Rounsevell}]{Arneth2014}
Arneth, A., Brown, C., Rounsevell, M.: Global models of human decision-making
  for land-based mitigation and adaptation assessment. Nature Climate Change
  \textbf{4}, 550--557 (2014), \doi{10.1038/nclimate2250}

\bibitem[{Axelrod(1997)}]{Axelrod1997}
Axelrod, R.: The dissemination of culture: A model with local convergence and
  global polarization. Journal of Conflict Resolution \textbf{41}(2), 203--226
  (1997), \doi{10.1177/0022002797041002001}

\bibitem[{Axtell and Farmer(2022)}]{Axtell2022}
Axtell, R., Farmer, J.: Agent-based modeling in economics and finance: Past,
  present, and future. Tech. Rep. 2022-10, Institute for New Economic Thinking
  at the Oxford Martin School (2022)

\bibitem[{Axtell(2016)}]{Axtell2016}
Axtell, R.L.: 120 million agents self-organize into 6 million firms: {A} model
  of the {U.S.} private sector. In: Proc.\ of AAMAS'16, pp. 806--816 (2016)

\bibitem[{Blandin et~al.(2017)Blandin, Colglazier, O'Hare, and
  Brenner}]{Blandin2017}
Blandin, N., Colglazier, C., O'Hare, J., Brenner, P.: Parallel python for
  agent-based modeling at a global scale. In: Proc.\ of CSSSA'17 (2017),
  \doi{10.1145/3145574.3145588}

\bibitem[{Blumofe et~al.(1995)Blumofe, Joerg, Kuszmaul, Leiserson, Randall, and
  Zhou}]{Blumofe1995}
Blumofe, R.D., Joerg, C.F., Kuszmaul, B.C., Leiserson, C.E., Randall, K.H.,
  Zhou, Y.: Cilk: An efficient multithreaded runtime system. In: Proc.\ of
  PPOPP'95, pp. 207--216 (1995), \doi{10.1145/209936.209958}

\bibitem[{Blythe and Tregubov(2018)}]{Blythe2018}
Blythe, J., Tregubov, A.: Farm: Architecture for distributed agent-based social
  simulations. In: Proc.\ of Second International Workshop on Massively
  Multiagent Systems, pp. 96--107 (2018)

\bibitem[{Breitwieser et~al.(2021)Breitwieser, Hesam, de~Montigny, Vavourakis,
  Iosif, Jennings, Kaiser, Manca, Meglio, Al-Ars, Rademakers, Mutlu, and
  Bauer}]{Breitwieser2021}
Breitwieser, L., Hesam, A., de~Montigny, J., Vavourakis, V., Iosif, A.,
  Jennings, J., Kaiser, M., Manca, M., Meglio, A.D., Al-Ars, Z., Rademakers,
  F., Mutlu, O., Bauer, R.: {BioDynaMo}: a modular platform for
  high-performance agent-based simulation. Bioinformatics \textbf{38}, 453--460
  (2021), \doi{10.1093/bioinformatics/btab649}

\bibitem[{ten Broeke et~al.(2021)ten Broeke, van Voorn, Ligtenberg, and
  Molenaar}]{DBLP:journals/jasss/BroekeVLM21}
ten Broeke, G., van Voorn, G., Ligtenberg, A., Molenaar, J.: The use of
  surrogate models to analyse agent-based models. J. Artif. Soc. Soc. Simul.
  \textbf{24}(2) (2021), \doi{10.18564/jasss.4530}

\bibitem[{Băbeanu et~al.(2018)Băbeanu, van~de Vis, and
  Garlaschelli}]{Babeanu2018}
Băbeanu, A.I., van~de Vis, J., Garlaschelli, D.: Ultrametricity increases the
  predictability of cultural dynamics. New Journal of Physics \textbf{20}(10),
  103026 (2018), \doi{10.1088/1367-2630/aae566}

\bibitem[{Dagum and Menon(1998)}]{dagum1998openmp}
Dagum, L., Menon, R.: {OpenMP}: an industry standard {API} for shared-memory
  programming. Computational Science \& Engineering, IEEE \textbf{5}(1), 46--55
  (1998)

\bibitem[{Filatova et~al.(2011)Filatova, Voinov, and van~der
  Veen}]{DBLP:journals/envsoft/FilatovaVV11}
Filatova, T., Voinov, A.A., van~der Veen, A.: Land market mechanisms for
  preservation of space for coastal ecosystems: An agent-based analysis.
  Environ. Model. Softw. \textbf{26}(2), 179--190 (2011),
  \doi{10.1016/j.envsoft.2010.08.001}

\bibitem[{Gemmetto et~al.(2014)Gemmetto, Barrat, and Cattuto}]{Gemmetto2014}
Gemmetto, V., Barrat, A., Cattuto, C.: Mitigation of infectious disease at
  school: targeted class closure vs school closure. BMC Infectious Diseases
  \textbf{14}(1), 695 (2014), \doi{10.1186/s12879-014-0695-9}

\bibitem[{Gomez et~al.(2021)Gomez, Prieto, Leon, and Rodríguez}]{Gomez2021}
Gomez, J., Prieto, J., Leon, E., Rodríguez, A.: Infekta -- an agent-based
  model for transmission of infectious diseases: The {COVID-19} case in
  {B}ogotá, {C}olombia. PLOS ONE \textbf{16}(2), 1--16 (2021),
  \doi{10.1371/journal.pone.0245787}

\bibitem[{Ham(2012)}]{Ham2012}
Ham, D.R.: Dynamic scheduling in multicore processors. Ph.D. thesis, The
  University of Manchester (United Kingdom) (2012)

\bibitem[{Lamperti et~al.(2018)Lamperti, Roventini, and Sani}]{Lamperti2018}
Lamperti, F., Roventini, A., Sani, A.: Agent-based model calibration using
  machine learning surrogates. Journal of Economic Dynamics and Control
  \textbf{90}, 366--389 (2018), \doi{10.1016/j.jedc.2018.03.011}

\bibitem[{Leeuw et~al.(2022)Leeuw, Ziabari, and
  Sharpanskykh}]{DBLP:conf/mabs/LeeuwZS22}
Leeuw, B.D., Ziabari, S.S.M., Sharpanskykh, A.: Surrogate modeling of
  agent-based airport terminal operations. In: Proc.\ of MABS'22, pp. 82--94
  (2022), \doi{10.1007/978-3-031-22947-3\_7}

\bibitem[{Parry(2020)}]{Parry2020}
Parry, H.R.: Agent-based modeling, large-scale simulations. In: Complex Social
  and Behavioral Systems: Game Theory and Agent-Based Models, pp. 913--926,
  Springer (2020), \doi{10.1007/978-1-0716-0368-0\_9}

\bibitem[{Plesser et~al.(2007)Plesser, Eppler, Morrison, Diesmann, and
  Gewaltig}]{Plesser2007}
Plesser, H.E., Eppler, J.M., Morrison, A., Diesmann, M., Gewaltig, M.O.:
  Efficient parallel simulation of large-scale neuronal networks on clusters of
  multiprocessor computers. In: Proc.\ of Euro-Par'07, pp. 672--681 (2007)

\bibitem[{Scheffer et~al.(1995)Scheffer, Baveco, DeAngelis, Rose, and van
  Nes}]{Scheffer1995}
Scheffer, M., Baveco, J., DeAngelis, D., Rose, K., van Nes, E.:
  Super-individuals: a simple solution for modelling large populations on an
  individual basis. Ecological Modelling \textbf{80}(2--3), 161--170 (1995)

\bibitem[{Scheutz and Harris(2010)}]{DBLP:series/sci/ScheutzH10}
Scheutz, M., Harris, J.: Adaptive scheduling algorithms for the dynamic
  distribution and parallel execution of spatial agent-based models. In:
  Parallel and Distributed Computational Intelligence, pp. 207--233, Springer
  (2010), \doi{10.1007/978-3-642-10675-0\_10}

\bibitem[{Scheutz and Schermerhorn(2006)}]{DBLP:journals/jpdc/ScheutzS06}
Scheutz, M., Schermerhorn, P.W.: Adaptive algorithms for the dynamic
  distribution and parallel execution of agent-based models. J.\@ Parallel
  Distributed Comput. \textbf{66}(8), 1037--1051 (2006),
  \doi{10.1016/j.jpdc.2005.09.004}

\bibitem[{Tregubov and Blythe(2020)}]{DBLP:conf/mabs/TregubovB20}
Tregubov, A., Blythe, J.: Optimization of large-scale agent-based simulations
  through automated abstraction and simplification. In: Proc.\ of MABS'20, pp.
  81--93 (2020), \doi{10.1007/978-3-030-66888-4\_7}

\end{thebibliography}

\end{document}